\documentclass[aps,prc,twocolumn,showpacs,preprintnumbers,amsmath,amssymb]{revtex4}
\usepackage{graphicx}% Include figure files
\usepackage{dcolumn}% Align table columns on decimal point
\usepackage{bm}% bold math
\usepackage{color}% color
\usepackage{float}

\begin{document}

\title{Fusion systematics for weakly bound nuclei using neutron flow and collective degrees of freedom}% Force line breaks with \\

\author{S. Appannababu$^1$\footnote{appannababu@gmail.com}}
\author{V. V. Parkar$^{2,3}$\footnote{vparkar@barc.gov.in}}
\author{V. Jha$^{2,3}$\footnote{vjha@barc.gov.in}}
\author{S. Kailas$^4$\footnote{swaminathankailas305@gmail.com}}
\affiliation{$^1$Department of Nuclear Physics, Andhra University, Visakhapatnam - 530003, India}
\affiliation{$^2$Nuclear Physics Division, Bhabha Atomic Research Centre, Mumbai - 400085, India}
\affiliation{$^3$Homi Bhabha National Institute, Anushaktinagar, Mumbai - 400094, India}
\affiliation{$^4$UM-DAE Centre for Excellence in Basic Sciences, Mumbai - 400098, India}

\date{\today}% It is always \today, today,
             %  but any date may be explicitly specified

\begin{abstract}

A systematic analysis of the fusion cross sections around the Coulomb barrier energies with  stable weakly bound ($^{6}$Li,$^{7}$Li,$^{9}$Be) and strongly bound $^{12}$C projectiles on various targets was performed by using neutron flow model and coupled channels approach. The analysis show that both the models are successful in explaining the near barrier fusion data. Further, it is also observed that the collective degrees of freedom as well as the neutron flow influence the near barrier fusion process involving weakly bound projectiles.
\end{abstract}

\maketitle

\section{Introduction}
Fusion of two nuclei at energies below the Coulomb barrier is a topic of great interest due to its relation in understanding nucleosynthesis of elements and probing underlying rich nuclear dynamics. It is well established that the fusion cross-sections are significantly enhanced at sub-barrier energies compared to the one dimensional barrier penetration model (1DBPM) \cite{Dasgupta1998, Becker88, Bala98}. The enhancement of the fusion cross-sections is explained due to coupling of relative motion with the various internal degrees of freedom  of the interacting nuclei \cite{Zag2003, Stephan1995, Bierman1996}. In addition to this, the role of neutron rearrangement process on the fusion cross-sections  has been also explored for various projectile-target systems \cite{Zag2003}. 

There are several theoretical prescriptions proposed to describe the enhancement in the fusion cross-sections at near barrier energies and it is observed that two different models are very much successful in explaining the enhancement in the cross-sections  \cite{Dasso83, Broglia1983, Stelson88, Stelson90}. The first model  is the Coupled Channels (CC) model, which is based on the effect of couplings of the incident projectile and the target which leads to lowering the barrier height of the interacting system. Because of reduction of barrier height, an enhancement of the fusion cross-section compared to the 1DBPM \cite{Dasso83, Broglia1983} is observed. The second model is the Stelson model, which is based on the neutron flow due to the exchange of neutrons between the interacting projectile and target nuclei \cite{Stelson88, Stelson90}. It is observed that both these models predict different barrier distributions for various reactions studied. In the past, there were only a few studies carried out in order to understand the enhancement of the sub barrier fusion cross-sections by using both the models \cite{Kailas93, VinodKumar96, Shapira93, Vandenbosch92} and make a comparison between these models. Systematic study of several reactions using Stelson model indicates that there is an early onset of free neutron flow  between the interacting projectile and the target at relatively larger inter nuclear distances.  This leads to an enhancement of the fusion cross-sections at below barrier energies and the correlation of barrier shifts for various systems can also be successfully explained by this model \cite{Shapira93, VinodKumar96}. On the other hand, the coupled channel approach is well established and widely used method of calculating sub-barrier fusion, which is very successful in explaining the experimental data for a variety of nuclear systems \cite{Kailas93}. 
There have been only a few attempts in literature, in order to make comparative study of these two methods and  identify which of these mechanisms is more appropriate to explain the experimental data on sub barrier fusion. In one of the studies, it was reported that the coupled channels method is better correlated with the experimental data, when compared to the neutron flow model \cite{Kailas93}. In the past, reactions involving strongly bound stable projectiles were investigated by using both the models discussed above. These investigations show that there is a very good correlation between the experimental data and the theoretical calculations \cite{Kailas93, VinodKumar96, Shapira93, Vandenbosch92}. It is very interesting to make similar study for the reactions involving weakly bound projectiles (WBP) and compare the results with the systematics of strongly bound projectile (SBP) on various targets. It is well known that for the reactions involving WBPs such as, $^{6}$Li,$^{7}$Li and $^{9}$Be, due to their small breakup thresholds, there is an enhancement in the fusion cross-sections at sub barrier energies. There are several reports in literature showing that the  reaction cross-sections involving WBPs were very high around the barrier energies \cite{Jha20, Canto06}. In this context, it is very much interesting to investigate the systematic of WBPs on various targets by using both the formalisms discussed above \cite{Kailas93}. 
 
In this paper, we investigate the systematics of WBPs on various targets by using both the coupled channel and Stelson model and compare the results with SBP $^{12}$C on various targets. The paper is organized as follows: In section II, the methodology for neutron flow based model and coupled channel models for calculation of fusion cross-sections are described. In section III, the results from WBP and SBP $^{12}$C projectile on various targets has been presented. The summary and conclusions are given in section IV.
 
\section{Methodology}
In this paper,  the complete fusion (CF) \color{black}cross-sections for $^{6}$Li,$^{7}$Li, $^{9}$Be and $^{12}$C projectiles on various targets was analyzed by using neutron flow model and coupled channels model. For completeness of the paper, a brief introduction about the two models was discussed here. The well known expression for the fusion cross-section for reactions with projectile energies greater than the barrier (B) can be written as  
\begin{equation}
\sigma_{fus} = \pi R_{b}^{2} (1 - \frac{B}{E})
\end{equation}
where B, R$_{b}$, E are the Coulomb barrier, Coulomb radius and energy in the c.m. frame respectively \cite{Frobrich96}. 
According to the Stelson model, at near barrier energies the fusion barriers can be explained by a flat distribution of barriers with a threshold energy cutoff (T$_{exp}$) \cite{Stelson88, Stelson90}. This barrier corresponds to the energy at which the interacting nuclei come sufficiently close to each other for neutrons to flow freely between target and projectile. So, the above expression transforms at near barrier energies to 
\begin{equation}
\sigma_{fus} = \pi R_{b}^{2} \frac{(E - T_{exp})^{2}}{4E(B - T_{exp})}
\end{equation}
The maximum value of the merged neutron potential V$_{max}$ can be calculated by assuming the neutron shell potential  centered on each of the interacting nuclei. In this configuration, the distance between the interacting nuclei is given by R$_{t}$, the distance at which the threshold barrier (T$_{cal}$) is reached. According to this model, if the merged neutron potential (V$_{max}$) is equal or lower than the binding energy of the valence neutron of the two interacting nuclei then only the neutron flow takes place. Further the extent of the barriers discussed above B - T$_{exp}$ are correlated by the difference between the Coulomb barrier and the threshold barrier B - T$_{cal}$.\\

In the coupled channel formalism, it is well established that the coupling between the incident projectile and target channels (vibrational, rotational, transfer) can modify the barrier heights \cite{Dasso83, Broglia1983}. By including the transmission probabilities and strength of the couplings (F) through the modified barriers, the fusion cross-section can be calculated. If only inelastic couplings are considered in the analyses, the channel couplings (F) may lead to the decrease or increase in the barrier height and it is expected that the B - T$_{exp}$ values will be related to F. The coupling strength for inelastic excitations to collective states can be  calculated from the deformation parameter $\beta_{\lambda,k}$, where $\lambda$ is the multi-polarity of the transition and k is the excited state of the nuclei (target or projectile) \cite{Kailas93}. \\

By using the above discussed formalism, complete fusion data of $^{6}$Li,$^{7}$Li, $^{9}$Be and $^{12}$C projectiles on various targets was analyzed.  The inelastic states of the target were considered in the coupled channel calculations for all the WBP ($^{6}$Li,$^{7}$Li, $^{9}$Be) and SBP ($^{12}$C) systems, while the first projectile inelastic state for $^{12}$C (4.4 MeV) and $^7$Li (0.48 MeV) were considered. In the case $^6$Li and $^9$Be, inelastic excitation corresponding to 2.18 MeV and 2.43 MeV resonance states were considered respectively in the coupled channel calculations. \color{black} Depending upon the correlation plots between B - T$_{exp}$ versus F and B - T$_{exp}$ versus B - T$_{cal}$, one can estimate which method is more reliable to explain the fusion cross-sections around the barrier energies. The values of B and $R_{b}$ are calculated by fitting the fusion data at above barrier energies by using the Eq.1 and the value of T$_{exp}$ was calculated by fitting the data at near barrier energies by using Eq.2. In order to calculate the effective value of the strength of the couplings (F), CCFULL code \cite{Hagi99} was used. Fusion cross section was calculated with and without coupling at deep sub-barrier energies and equated the ratio to exp(F.$\epsilon$) where $\epsilon$ is the barrier curvature. Knowing $\epsilon$, F can be determined. The deformation, multipolarity and transition strengths of the excited states of the projectile/ target used in the CCFULL code are taken from the literature  \cite{Raman87,Kibedi02,Pritichenkov16}. Previously, it was shown that at sub-barrier energies, this formalism is very much valid in order to extract the effective value of F \cite{Kailas93, Landowne84}.\\

B - T$_{cal}$ values were calculated by the following procedure. The average one/two neutron separation energies for various projectiles and targets were taken from the mass table \cite{Wapstra88}. As the S$_{n}$ values of WBPs $^{6,7}$Li and $^{9}$Be are smaller than those of the targets, the neutron flow  is from the projectile to the target. However, in the case of $^{12}$C as the S$_{n}$ value is nearly 19 MeV, the neutron flow is from the target to the projectile. By using the neutron potential values given in Ref. \cite{Stelson90} for both the target and the projectile, the interacting distance R$_{t}$ between the two has been optimized such that  the merged neutron potential V$_{max}$ at this distance is equal to the S$_{n}$ value of the target or projectile (whichever is lower). The T$_{cal}$  value has been computed using  R$_{t}$ as discussed in Ref. \cite{Kailas93}. The merged neutron potentials for the two reactions $^{12}$C + $^{208}$Pb and $^{9}$Be + $^{208}$Pb calculated are shown in Fig.1. One can observe that, the V$_{max}$  values match with the S$_{n}$ values of $^{208}$Pb and $^{9}$Be respectively.  Here, we have used S$_{2n/2}$ for the targets to take care of the odd – even effects \cite{Kailas93}. 
\begin{figure}
\centering
\includegraphics[width=85 mm]{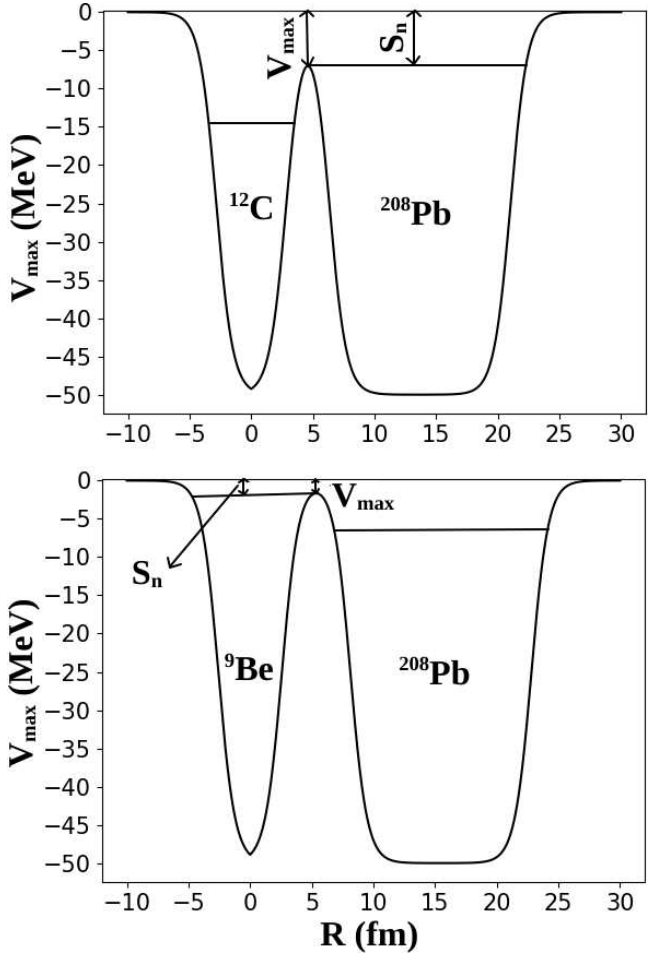}
\caption{ Merged neutron potentials at R$_{t}$, for the reactions $^{12}$C + $^{208}$Pb (upper panel) and $^{9}$Be + $^{208}$Pb  (lower panel).}
\label{fig:Fig1a}
\end{figure}
\section{Results and Discussion}
By using the above discussed methodologies, the different parameters B, R$_{b}$, T$_{exp}$, S$_{n}$, R$_{t}$,T$_{cal}$ and F were extracted for $^{12}$C, $^{6,7}$Li and $^{9}$Be projectiles on various targets. The different parameters calculated by using both the methodologies are given in Table\ \ref{tab-1}. 
\begin{table*}
\begin{center}\caption{Summary of near barrier fusion analysis for $^{12}$C, $^{6,7}$Li and $^{9}$Be projectiles on various targets.}
\label{tab-1}
\begin{tabular}{|c|c|c|c|c|c|c|c|c|c|c|}
\hline
 Reaction  &  B  & R$_{0}$ &  T$_{exp}$ (err.)  &  S$_{2n/2}$  & R$_{t}$  & T$_{cal}$ & B - T$_{cal}$ & B - T$_{exp}$ (err.) & F  & Refs.\\
  & (MeV) & (fm) & (MeV)  & S$_{n}$ (MeV)  & (fm)  &(MeV)&  (MeV) &  (MeV)& (MeV) & \\
\hline
 $^{12}$C +  $^{46}$Ti   & 21.09 (0.47) & 9.65  & 18.57 (0.10) & 11.36 & 10.07  & 18.87 & 2.22 & 2.52 (0.48)& 0.34  &  \cite{Bozek86} \\
 $^{12}$C +  $^{48}$Ti   & 20.32 (0.66) & 7.83  & 17.46 (0.22) & 10.25 & 10.29  & 18.46 & 1.86 & 2.86 (0.70)& 0.32  &  \cite{Bozek86} \\
 $^{12}$C +  $^{50}$Ti   & 19.32 (1.09) & 7.68  & 17.76 (0.16) & 9.54 & 10.46  & 18.16 & 1.16 & 1.56 (1.10)& 0.27  &  \cite{Bozek86} \\ 
   $^{12}$C +  $^{92}$Zr   & 31.90 (0.14) & 9.13  & 29.66 (0.02) & 7.91 & 11.65  & 29.66 & 2.28 & 2.24 (0.14)& 0.41  &  \cite{Newton01} \\
  $^{12}$C  +  $^{144}$Sm  & 45.98 (0.46) & 10.49 &44.23 (0.08)  & 9.56 & 12.39  & 43.22  &2.75 & 1.75 (0.47) & 0.47 & \cite{Abriola92} \\
    $^{12}$C +  $^{152}$Sm  & 47.27 (0.92) &11.30 &42.0 (0.19)  & 6.92 & 12.99  & 41.24 & 6.03 & 5.27 (0.94) & 1.54 &\cite{Brod75} \\
    $^{12}$C +  $^{181}$Ta  & 52.97 (1.17) & 10.83 & 48.62 (0.1) & 7.11 & 13.35  & 47.26 & 5.17 &4.35 (1.17) & 0.71   &\cite{Crippa94} \\
  $^{12}$C +  $^{194}$Pt  & 54.99 (0.51) & 10.54 & 52.35 (0.04)& 7.30 & 13.47  & 50.03 & 4.96 & 2.64 (0.51) & 0.64 & \cite{Aradhana01} \\
  $^{12}$C +  $^{198}$Pt   & 54.86 (0.34) & 10.32 & 52.76 (0.02)& 6.70 & 13.64  & 49.39 & 5.47 & 2.1 (0.34) & 0.81 &\cite{Aradhana01} \\
  $^{12}$C  +  $^{204}$Pb  & 54.15 (1.59) & 9.85 &51.37(0.33) & 7.65 & 13.52  & 52.39 & 1.76 & 2.78 (1.62) & 0.64  &\cite{Sagaidak03} \\
  $^{12}$C +  $^{206}$Pb  & 56.75 (2.29)& 11.20 &53.19 (0.59)& 7.41 & 13.59  & 52.11 & 4.64 & 3.56 (2.37)& 0.65 &\cite{Sagaidak03}\\
  $^{12}$C +  $^{208}$Pb  & 55.47 (0.69) & 9.04 &54.17 (0.06) & 7.05 & 13.69  & 51.75 & 3.72 &1.3 (0.50)&0.64  &\cite{Mukherjee07}\\
%  $^{12}$C +  $^{209}$Bi  & 59.04 & 57.13  & 6.27 & 1.91 (0.31) & 0.178  & 15.07 & \cite{Ge07}\\
\hline   
%  \hline
    $^{6}$Li  +  $^{28}$Si  & 5.73 (0.7)) & 6.22 &5.01 (0.12)  & 5.66 & 9.84  & 5.27 & 0.46 & 0.72 (0.7) & 0.19 & \cite{Mandira10}\\
    $^{6}$Li  +  $^{64}$Ni  & 11.27 (0.9)) & 7.81 &9.74 (0.06)  &  & 10.88  & 11.12 & 0.15 & 1.53 (0.9) & 0.29 & \cite{Moin14}\\
    $^{6}$Li  +  $^{90}$Zr  & 18.46 (0.3) & 8.24 &15.14 (0.07)  &  & 11.63  & 14.85 & 3.61 & 3.32 (0.3) & 0.31 & \cite{Kumawat12}\\
    $^{6}$Li  +  $^{124}$Sn  & 20.84 (0.4)) & 8.67 &17.08 (0.11)  &  & 12.26  & 17.62 & 3.22 & 3.76 (0.41) & 0.35 & \cite{VVP18b}\\
    $^{6}$Li  +  $^{144}$Sm  & 26.03 (0.7)) & 7.43 &22.71 (0.03)  &  & 12.58  & 21.29 & 4.73 & 3.32 (0.7) & 0.33 & \cite{Rath09}\\
  $^{6}$Li +  $^{152}$Sm   & 25.71 (0.59)& 8.24 & 21.75 (0.04)  & & 12.69  & 21.09 & 4.61 & 3.96 (0.6) & 0.33 &\cite{Rath12} \\
    $^{6}$Li +  $^{198}$Pt  & 28.2 (1.01)& 8.71 & 24.3 (0.16)  &  & 13.30 & 25.32 & 2.87 & 3.9 (1.02) & 0.37 & \cite{Shrivastava09}\\
    $^{6}$Li +  $^{197}$Au  & 28.42 (0.04)& 8.39 & 25.2 (0.09) &  & 13.29 & 27.67 & 0.74 & 3.2 (0.1)& 0.34 & \cite{Pals14} \\
  $^{6}$Li +  $^{208}$Pb  & 27.72 (1.07)& 7.41 & 26.58 (0.09) &  & 13.42 & 26.39 & 1.33 & 1.14 (1.1) & 0.35 &  \cite{Wu03}\\
  $^{6}$Li +  $^{209}$Bi   & 29.96 (0.83)& 8.78 & 26.57 (0.10)&  & 13.44  & 28.69 & 1.27 & 3.39 (0.84) & 0.35 & \cite{Dasgupta02} \\
  
\hline   
%  \hline
   $^{7}$Li  +  $^{59}$Co  & 11.61 (0.26)& 7.64 & 9.44 (0.1)  & 7.25 & 10.66  & 10.94 & 0.68 & 2.17 (0.3)& 0.24 & \cite{Beck03} \\
    $^{7}$Li  +  $^{124}$Sn  & 21.46 (0.18) & 8.92 & 17.8 (0.03)  & & 12.02  & 17.97 & 3.49 & 3.66 (0.18) & 0.29 & \cite{VVP18}\\
   $^{7}$Li  +  $^{144}$Sm  & 24.82 (0.15)& 8.71 & 21.7 (0.06)  &  & 12.28  & 21.81 & 3.01 & 3.12 (0.16)& 0.31 & \cite{Rath13} \\
  $^{7}$Li +  $^{152}$Sm   & 24.18 (0.13)& 8.50 & 20.89 (0.04)  &  & 12.45  & 21.50 & 2.67 & 3.29 (0.13) & 0.40 &\cite{Rath13}\\
    $^{7}$Li +  $^{198}$Pt  & 28.81 (0.45)& 9.71 & 26.09 (0.07)  &  & 13.06 & 25.79 & 3.02 & 2.71 (0.45) & 0.32 & \cite{Ara13} \\
    $^{7}$Li +  $^{197}$Au  & 28.45 (0.99)& 9.96 & 26.09 (0.34) &  & 13.05 & 26.15 & 2.30 & 2.36 (1.05)& 0.31 &\cite{Pals14} \\
  $^{7}$Li +  $^{209}$Bi  & 29.52 (0.67)& 9.62 & 26.56 (0.10) &  & 13.20  & 27.18 & 2.34 & 2.96 (0.68) & 0.33 & \cite{Dasgupta02} \\
 
\hline   
  $^{9}$Be +  $^{89}$Y   & 22.49 (0.46)& 7.78 & 20.77 (0.08)  & 1.66 & 13.66  & 16.44 & 6.05 & 1.72 (0.47) & 0.38 & \cite{Pals10}\\
  $^{9}$Be  +  $^{124}$Sn  & 26.97 (0.26)& 11.29 & 24.11 (0.04)  &  & 14.31  & 20.13 & 6.84 & 2.86 (0.26)& 0.29 &\cite{Parkar10} \\
  $^{9}$Be +  $^{144}$Sm  & 32.13 (1.38)& 9.43 & 28.75 (0.17)  &  & 14.63  & 24.42 & 7.71& 3.38 (1.39) & 0.41 & \cite{Gomes2006} \\
  $^{9}$Be +  $^{208}$Pb  & 40.19 (0.23)& 9.72 & 36.69 (0.05) &  & 15.47  & 30.53 & 9.66 & 3.5 (0.24) & 0.51 &  \cite{Dasgupta04} \\
  $^{9}$Be +  $^{209}$Bi  & 37.84 (0.94)& 7.29 & 35.23 (0.06) &  & 15.48 & 30.88 & 6.96 & 2.61 (0.94)& 0.49 &  \cite{Dasgupta04}\\

\hline   
\end{tabular}
\end{center}
\end{table*}

As a typical example, the fusion cross section data for the reaction $^{9}$Be + $^{208}$Pb \cite{Dasgupta04} along with the Stelson model calculations are shown in Fig.\ \ref{fig:Fig2}. 
\begin{figure} 
\centering
\includegraphics[width=80 mm]{Fig2.eps}
\caption{Fusion cross section data available for $^{9}$Be + $^{208}$Pb system \cite{Dasgupta04} along with the Stelson model calculations. Continuous and dashed lines are Stelson model fits from the Equations 1 and 2, respectively.}
\label{fig:Fig2}
\end{figure}
\begin{figure} 
\centering
\includegraphics[width=90 mm]{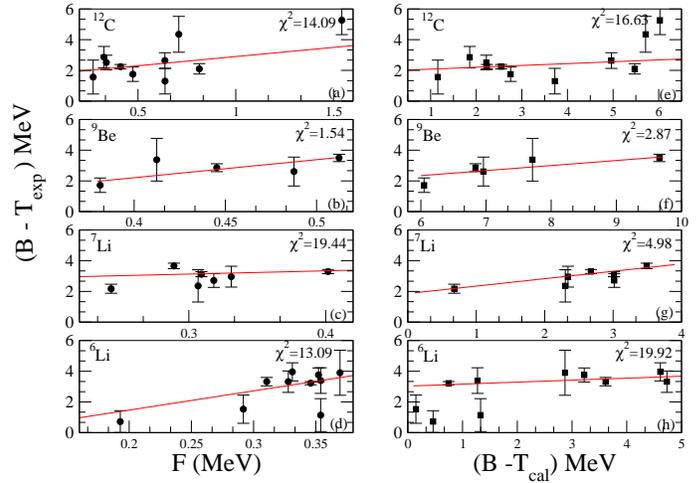}
\caption{The extracted values of B - T$_{exp}$ as a function of F (left panel) and  B - T$_{cal}$ (right panel) for the strongly bound $^{12}$C and weakly bound $^{6,7}$Li and $^{9}$Be projectiles on various targets \cite{Bozek86,Newton01,Abriola92,Brod75,Crippa94,Aradhana01,Sagaidak03,Mukherjee07,Mandira10,Moin14,Kumawat12,VVP18b,Rath09,Rath12,Shrivastava09,Pals14,Wu03,Beck03,VVP18,Dasgupta02,Rath13,Ara13,Pals10,Parkar10,Gomes2006,Dasgupta04}. The lines are the best linear fits to the data.}
\label{fig:Fig3}
\end{figure}
\begin{figure}[htbp]
\centering
\includegraphics[width=90 mm]{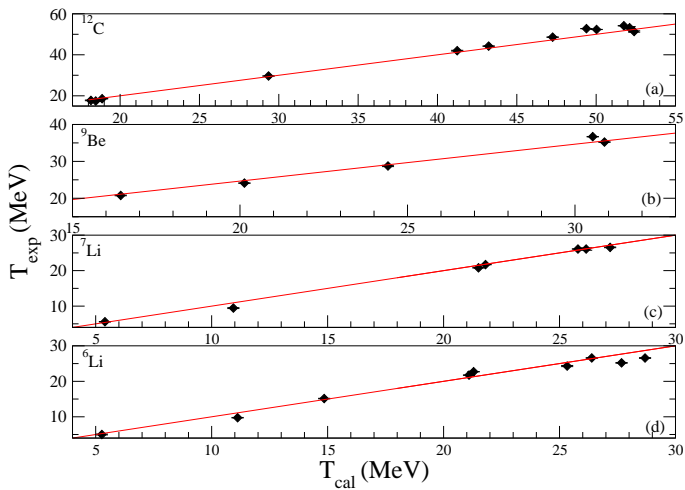}
\caption{The extracted values of T$_{exp}$ values as a function of T$_{cal}$ for the (a) strongly bound $^{12}$C and weakly bound (b) $^{9}$Be, (c) $^7$Li, and (d) $^6$Li projectiles on various targets. Lines are fit to the data (See text for details). \cite{Bozek86,Newton01,Abriola92,Brod75,Crippa94,Aradhana01,Sagaidak03,Mukherjee07,Mandira10,Moin14,Kumawat12,VVP18b,Rath09,Rath12,Shrivastava09,Pals14,Wu03,Beck03,VVP18,Dasgupta02,Rath13,Ara13,Pals10,Parkar10,Gomes2006,Dasgupta04}.}
\label{fig:Fig4}
\end{figure}

The values of the B - T$_{exp}$ as a function of F (left panel) and B - T$_{cal}$ (right panel) are plotted in Fig.\ \ref{fig:Fig3}. From the present analysis (Fig.3), it can be observed that for the reactions induced by $^{12}$C on various targets, the B - T$_{exp}$ values on the average increase with increasing F and  the B - T$_{exp}$ values also increase with increasing  B - T$_{cal}$. The present results are very much similar to the correlation results for reactions induced by $^{16}$O projectiles on various targets \cite{Kailas93}. It can also be observed that the correlation between B - T$_{exp}$ vs F is very much similar when compared to the correlation between  B - T$_{exp}$ vs B -T$_{cal}$. Further, Fig.3 shows that there is a strong correlation for the data analyzed by using the two models, since the results are well reproduced by a simple linear fit. Followed by these results, the experimental data for the reactions induced by WBPs  on various targets was analyzed in a similar way as discussed above for SBP $^{12}$C. From Fig.3, it is very interesting to observe a similar trend for WBPs as observed for $^{12}$C projectile on various targets. In case of WBPs also, one can observe that there is a good correlation between B - T$_{exp}$ vs B - T$_{cal}$ and F and all the analyzed experimental data is very well reproduced by linear fits. The $\chi^{2}$ fits for the two models are similar.  While both the Coupled Channels calculations and Stelson model describe the fusion data for a large  number of systems well, there is an important difference between the two models. As per the Stelson model, for all isotopes of a given Z the S$_n$ values are always lowest for the heaviest isotope, giving lowest T values for it irrespective of collectivity. In contrast, in the case of coupled channel effects due to collectivity, the isotope which is more collective has larger F value leading to lower T values. We have tried to look for this feature also in our data. However, as the B - T$_{exp}$ has large error bars and therefore we are not in a position to confirm the above expectation from our analysis. \color{black}

To understand the neutron transfer mechanism in Stelson model, further the T$_{cal}$ values for $^{12}$C, $^{6,7}$Li and $^{9}$Be projectiles on various targets as a function of T$_{exp}$ have been plotted in Fig. 4 which shows a very good correlation between T$_{exp}$ and T$_{cal}$ values.  In general, the fusion cross section at near and below  barrier energies is enhanced (for both strongly and weakly bound projectiles) when compared to the 1DBPM. Further, there is a small reduction at well above barrier for strongly bound projectiles, while there is a significant reduction  10 to 30 \%  in the  case of weakly bound projectiles. The behavior of fusion excitation function is understood in terms of coupling to various degrees of freedom like collective, transfer and breakup, where the latter two processes are more important for weakly bound nuclei. Diffuseness parameter has been varied by 20 \% to see its effect on the  R$_t$ and T$_{cal}$ values. The change in T$_{cal}$ values with diffuseness parameter varying  from 0.65 to 0.8 fm is around 1 MeV. Further, it is observed that a 20 \% change in fusion cross section changes the barrier  radius by about 10 \% and negligible change of the corresponding fusion barrier. Therefore, B-T values are not affected significantly. \color{black}

In order to understand the correlation between  T$_{exp}$ and T$_{cal}$, the experimental data has been fitted by using linear function. From different kinds of linear fits, one can conclude that for the reactions induced by $^{12}$C, $^{6,7}$Li projectiles on various targets a simple linear fit function y = x is sufficient to show a very good correlation between the calculated and experimental results. But for the reactions induced by $^{9}$Be projectile on various targets a simple linear fit is not sufficient to explain the analyzed data. Good correlation between  T$_{exp}$ and T$_{cal}$ is obtained by using the function y = x + c, where c is a constant. From this analysis, we can clearly conclude that T$_{cal}$ values are in good agreement with the values of  T$_{exp}$ for the reactions induced by $^{12}$C, $^{6,7}$Li projectile on various targets and for the reactions induced by $^{9}$Be projectiles on various targets T$_{exp}$ values are higher by few MeV when compared to the values of  T$_{cal}$. We need to understand, why this anomaly/different behavior  exists for weakly bound $^{9}$Be projectile on various targets, when compared to the other reactions. Further, from this analysis it can be concluded that in the case of strongly bound stable projectile $^{12}$C and weakly bound $^{6,7}$Li, $^{9}$Be, both the models (Stelson model and coupled channels formalism) are very much successful in explaining the near barrier fusion data. However from the neutron transfer model, one has to understand why the T$_{exp}$ values are few MeV higher than the T$_{cal}$ values for weakly bound $^{9}$Be projectiles.

From this analysis, it can be observed that both these models (Stelson and coupled channels) describe the trend of B - T$_{exp}$ variation successfully for strongly bound and weakly bound projectiles, indicating the validity of these models for sub barrier fusion data.  In the Coupled Channels calculations, the excited states of the targets have been included. In the case of $^{6}$Li and $^{9}$Be, the resonance states have been included as discussed above. Stelson neutron flow model considers only the transfer of neutrons between  the interacting nuclei at distances close to the grazing distance. However, the weak or strong binding of the projectile is taken into account is through the neutron binding energy of the interacting nuclei. \color{black}

\section{Summary and Conclusions}
In the present paper, the fusion data for weakly bound projectiles ($^{6,7}$Li and $^{9}$Be) and strongly bound projectile $^{12}$C around the coulomb barrier energies has been analyzed on various targets. The present results suggest that both the neutron flow and coupled channels models are successful in explaining the near barrier fusion data. Further from the Stelson model,  it has been observed that there is a good correlation between T$_{exp}$ and T$_{cal}$. For the reactions induced by $^{9}$Be projectiles on various targets, the T$_{exp}$ values are a few MeV higher than the corresponding T$_{cal}$ values and this feature needs to be investigated. The present results show that for reactions induced by WBPs on various targets, the collective degrees of freedom as well as the neutron flow are important in  influencing the  near barrier fusion phenomenon. In the case of Sm isotopes considered in the present work, while the S$_{n}$ (and hence the T values) values decrease with increase of the mass number of the isotopes,  the collectivity increases with increase of mass number. Hence B-T ( Stelson model) and F (collectivity) values increase with increase of mass number and we can not choose as to which out of the above two models is more appropriate to describe the data. However, if a series of isotopes were choosen, which have both S$_{n}$ and collectivity decrease with increase of mass number of the target, then it will be possible to select  between these two models. In this case,  if collectivity is the dominant mechanism, then B-T determined will  increase with decrease of  mass number as lower mass number has higher collectivity when compared to the heavier.  If neutron flow is more important mechanism, then B-T will increase with increase of mass number as the heavier isotope has a lower value of S$_{n}$ and hence T. It will be interesting to extend these studies to other WBPs and radioactive ions for a range of isotopes of  a given target.

\section{Acknowledgments}
The authors SK acknowledges INSA, India for the senior scientist fellowship and SA acknowledges the financial support from UGC, India.

% \smallskip
%\bibliography{reference}

\end{document}